\documentclass[a4paper, 12pt]{article}

\usepackage{graphicx}
\usepackage{url}
\usepackage{amsbsy,amssymb} 
\usepackage{amsmath}
\usepackage{abstract}
\usepackage{hyperref}
\usepackage{color}
\usepackage{float}
\usepackage[caption = false]{subfig}

\usepackage{enumitem}

\newtheorem{Proposition}{Proposition}
\newtheorem{Corollary}{Corollary}

\newcommand{\erfc}{\mathrm{erfc}\,}

\begin{document}

\title{Relativistic Voigt profile for unstable particles in high energy physics}
\author{Rados\l aw A. Kycia$^{1,2,a}$, Stanis\l aw Jadach$^{3,b}$}
\maketitle

\medskip
\small{
\centerline{$^{1}$Cracow University of Technology, Faculty of Physics,}
\centerline{Mathematics and Computer Science, PL-31155, Krak\'{o}w, Poland}

\medskip
\centerline{$^{2}$The Faculty of Science, Masaryk University,}
\centerline{Kotláøská 2, 602 00 Brno, Czechia}

\medskip
\small{
\centerline{$^{3}$The Henryk Niewodnicza\'nski Institute of Nuclear Physics}
\centerline{Polish Academy of Sciences}
\centerline{ul.\ Radzikowskiego 152, 31-342 Krak\'ow, Poland}
\bigskip

\centerline{$^{a}${\tt
kycia.radoslaw@gmail.com}, $^{b}${\tt stanislaw.jadach@ifj.edu.pl}}

\begin{abstract}
The Voigt profile is one of the most used special function in 
optical spectroscopy. In particle physics a version of this 
profile which originates from relativistic Breit-Wigner resonance distribution often appears, however, in the literature there is no strict derivation of its properties. 
The purpose of this paper is to define properly relativistic Voigt profile and describe it along the same way as their standard nonrelativistic counterparts.
\end{abstract}

%%%%%%%%%%%%%%%%%%%%%%%%%%%%%%%%%%%%%%%%%%%%%%%%%%%%%%%%%%%%%%%%%%%%%%%%%%
\section{Introduction}
%%%%%%%%%%%%%%%%%%%%%%%%%%%%%%%%%%%%%%%%%%%%%%%%%%%%%%%%%%%%%%%%%%%%%%%%%%
In nonrelativistic particle physics the cross section of an unstable 
particles/resonances is described by the (nonrelativistic) Breit-Wigner distribution \cite{Pilkuhn}
\begin{equation}
 \sigma_{0}(E;\mu,\Gamma):=\frac{\Gamma}{2\pi}\frac{1}
 {(E-\mu)^{2}+\left(\frac{\Gamma}{2}\right)^{2}},
 \label{Breit-Wigner}
\end{equation}
where the real variable $E>0$ is an energy of the resonance, the other real parameters 
are $\mu>0$ - the mass of unstable particle and $\Gamma>0$ - the 
width of the resonance. This distribution in mathematical applications is 
called the Cauchy distribution, when normalized to unity. Relativistic counterpart to 
(\ref{Breit-Wigner}) is also called (relativistic) Breit-Wigner 
distribution and is of the form \cite{Pilkuhn}
\begin{equation}
 \sigma_{2}(E;\mu,\Gamma):=\frac{\mu\Gamma}{\pi}\frac{1}
 {(E^{2}-\mu^{2})^{2}+(\mu\Gamma)^{2}}.
 \label{RelativisticBreitWigner}
\end{equation}

In measurement of resonances obeying these distributions, the additional factor which has 
to be taken into account is a beam broadening effect or detector 
sensitivity function, which is usually of the form of the Gauss distribution 
\begin{equation}
 G(E-E_{0};\sigma):=\frac{1}{\sigma\sqrt{2\pi}}e^{-\frac{(E-
 E_{0})^{2}}{2\sigma^{2}}},
 \label{GaussianPDF}
\end{equation}
where two real parameters have the following meaning - 
$E_{0}$(positive real number) is the position of the accelerator beam energy maximum and 
$\sigma>0$ - the dispersion of the beam or detector sensitivity 
around the maximum. The distribution which results from the original 
Breit-Wigner distribution and takes into account broadening effect of the Gaussian 
distribution has the form of a convolution integral
\begin{equation}
 V_{i}(E;\mu,\Gamma,\sigma):= \sigma_{i}*G = \int_{-\infty}^{\infty} 
 \sigma_{i}(E';\mu,\Gamma) G(E- E';\sigma) dE', \quad i=0,2.
 \label{convolution}
\end{equation}
The resulting distribution for $i=0$ (nonrelativistic case) is called the 
Voigt distribution (see section 7.19 of \cite{NIST}) and its standard 
normal form is presented in the Appendix \ref{Appendix_Voigt}. In the 
next part we define in the similar way, as we call it, the relativistic 
Voigt function which is (\ref{convolution}) for $i=2$ (relativistic 
case). This function is often used in High Energy Physics and, sadly 
to say, to our best knowledge was not extracted as a mathematical 
entity on its own, e.g., see excellent compendium of special functions \cite{NIST}. This paper is intended to fill this gap.

In the next section we define relativistic Voigt profile, outline its 
basic properties and define also duping function for the Voigt 
functions. In the Appendices technical details and derivations of these 
results are provided. Appendix \ref{Appendix_Voigt} summarizes well-known properties of standard Voigt profile.

%%%%%%%%%%%%%%%%%%%%%%%%%%%%%%%%%%%%%%%%%%%%%%%%%%%%%%%%%%%%%%%%%%%%%%%%%%%%%
\section{Relativistic Voigt profile}
Relativistic Voigt distribution can be introduced in the similar 
manner as for classical Voigt profile summarized in in Appendix \ref{Appendix_Voigt}. It is useful in considerations of relativistic unstable particles like the Higgs boson \cite{HiggsLineshape}.
The equation (\ref{convolution}) for $i=2$ is of the form
\begin{equation} 
 V_{2}(E;\mu,\Gamma,\sigma)=\int_{-\infty}^{\infty} \frac{\mu\Gamma}
 {\pi}\frac{1}{(E'^{2}-\mu^{2})^{2}+(\mu\Gamma)^{2}}\frac{1}
 {\sigma\sqrt{2\pi}}e^{-\frac{(E'-E)^{2}}{2\sigma^{2}}} dE'.
 \label{V2}
\end{equation}
Introducing the variable
\begin{equation}
 t = \frac{E-E'}{\sqrt{2}\sigma},
\end{equation}
we obtain
\begin{equation}
 V_{2}(E;\mu,\Gamma,\sigma)=\frac{\mu\Gamma}{\pi\sigma 
 \sqrt{2\pi}}\int_{-\infty}^{\infty} \frac{e^{-t^{2}}}
 {(E-\sqrt{2}\sigma t-\mu)^{2}(E-\sqrt{2}\sigma 
 t-\mu)^{2}+(\mu\Gamma)^{2}}\sigma\sqrt{2} dt.
\end{equation}
Defining new variables
\begin{equation}
 u_{1}:=\frac{E-\mu}{\sqrt{2}\sigma},\quad u_{2}:=\frac{E+\mu}
 {\sqrt{2}\sigma},\quad a := \frac{\Gamma\mu}{2\sigma^{2}},
\end{equation}
the distribution can be rewritten in the form
\begin{equation}
 V_{2}(E;\mu,\Gamma,\sigma)=\frac{1}{\sigma^{2} 2\sqrt{\pi}} \frac{a}
 {\pi}\int_{-\infty}^{\infty} \frac{e^{-t^{2}}}{(u_{1}-t)^{2}(u_{2}-
 t)^{2}+a^{2}} dt.
 \label{normalizedRelativisticVoigtProfile}
\end{equation}
New function which can be called relativistic line 
broadening function (it is an analogy to (\ref{line_broadening_function}) presented below) can be defined as
\begin{equation}
 H_{2}(a,u_{1},u_{2}):=\frac{a}{\pi}\int_{-\infty}^{\infty} 
 \frac{e^{-t^{2}}}{(u_{1}-t)^{2}(u_{2}-t)^{2}+a^{2}} dt,
 \label{relativistic_line_broadening_function}
\end{equation}
and (\ref{normalizedRelativisticVoigtProfile}) can be rewritten in the following form
\begin{equation}
 V_{2}(E;\mu,\Gamma,\sigma)=\frac{H_{2}(a,u_{1},u_{2})}{\sigma^{2} 
 2\sqrt{\pi}}.
 \label{V2H2correspondence}
\end{equation}
The relativistic $H_{2}$ depends on three parameters, not on two as 
nonrelativistic counterpart (\ref{line_broadening_function})

Let us give some simple properties of $H_{2}$. 
\begin{Proposition}
\label{Proposition:reflection}
We have
\begin{equation}
 \begin{array}{c}
  H_{2}(a,u_{1},u_{2})=H_{2}(a,u_{2},u_{1}) \\
  H_{2}(a,-u_{1},u_{2})=H_{2}(a,u_{1},-u_{2})\\
  H_{2}(-a,u_{1},u_{2})=-H_{2}(a,u_{1},u_{2}),
 \end{array}
 \label{Symmetry}
\end{equation}
from which results that 
\begin{equation}
 H_{2}(a,-u_{1},-u_{2})=H_{2}(a,u_{1},u_{2}).
\end{equation}
\end{Proposition}
These symmetry properties allows one to consider the 
function in the first octant $\{ a>0, u_{1}>0,u_{2}>0\}$ and then 
propagate the values to other octants.
%%%%%

From the Appendix \ref{Appendix_Normalization} and \ref{Appendix_u1u2large} it results that 
\begin{Proposition}
\label{Proposition:a0limit}
For $u_{1}\neq u_{2}$
\begin{equation}
 lim_{a\rightarrow 0^{\pm}}H_{2}(a,u_{1},u_{2})=\frac{\pm}{|u_{1}-
 u_{2}|}(e^{-u_{1}^{2}}+e^{-u_{2}^{2}}).
 \label{H2_limit_a_0}
\end{equation} 

For $u_{1}=u_{2}$ and nonzero $a$ the $H_{2}$ function is finite and 
when $a\rightarrow 0$ it becomes unbounded with assymptotics
\begin{equation}
 H_{2}(a,u_{1}=u,u_{2}=u)=\frac{e^{-u^2}}{\sqrt{2a}}+ \frac{e^{-u^{2}}}{\sqrt{2}}(2u^{2}-1)\sqrt{a} + O(a).
\end{equation}

When $|u_{1}|$ and $|u_{2}|$ tends to 
infinity for $a\neq 0$ then as the integrand in 
(\ref{relativistic_line_broadening_function}) tends to zero, so
$H_{2}$ also vanishes. It is also true when $a \rightarrow 0^{\pm}$ 
and $|u_{1} - u_{2}| \rightarrow \infty$. 

\end{Proposition}
%%%

Summing up, $H_{2}$ for fixed $u_{1}\neq u_{2}$ is discontinuous and 
singular for $u_{1}=u_{2}$ when $a\rightarrow 0$. In Fig. 
\ref{Fig:H2} there is a plot of a few examples of the $H_{2}$ function.
%%%%%%%%%%%%%%%%%%%%%%%%%%%%%%%%%%%%%%%%%%%%%%%
\begin{figure}[!htb]
\centering
    \includegraphics[width=0.8\textwidth]{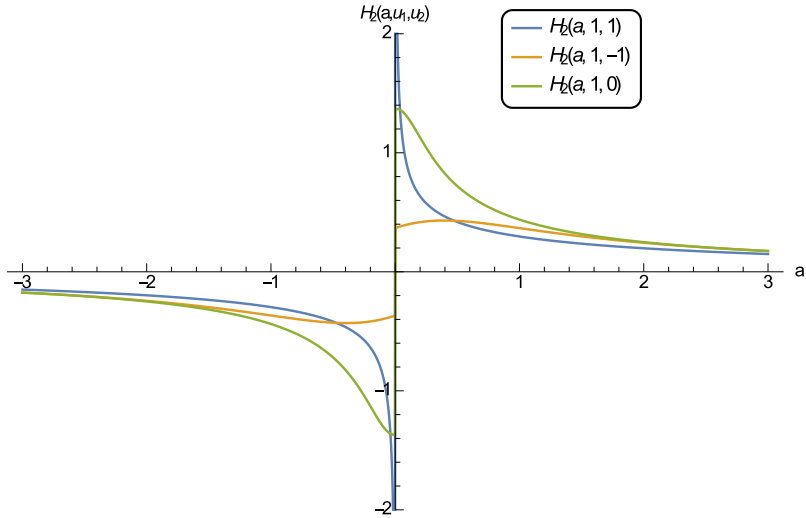}
    \label{Fig:H2}
    \caption{The relativistic Voigt distribution 
    (\ref{relativistic_line_broadening_function}) for a few values of 
    parameters. The values at $a\rightarrow 0$, according to 
    (\ref{H2_limit_a_0}) are  
    for $H_{2}(a,1,0)$:  $1+\frac{1}{e}$ and for                                                                                                                                                                                                                                                                        
    $H_{2}(a,1,-1)$: $\frac{1}{e}$. The function $H_{2}(a,1,1)$ at 
    $a=0$ is unbounded.}
\end{figure}
%%%%%%%%%%%%%%%%%%%%%%%%%%%%%%%%%%%%%%%%%%%%%%%%

The limit $a\rightarrow 0^{\pm}$ of $H_{2}$ in Proposition \ref{Proposition:a0limit} results from the following representation that is derived in Appendix \ref{Appendix_Normalization}, namely,
%%%%%%%%%%%%%%%%%%%%%%%%%%%%%%%%%%%%%%%%%%%%%%%%%
\begin{Proposition}
\label{Proposition:H2ErfcRepresentation}
$H_{2}(a,u_{1},u_{2})$ for $a>0$ can be represented by the complementary error function (Chapter 7 of \cite{NIST}) in the form
\begin{equation}
\begin{array}{c}
 H_{2}(a, u_{1}, u_{2}) = \frac{1}{2w_{1}}\left( e^{-(t_{1+})^{2}} \erfc(-it_{1+}) + e^{-(t_{1-})^{2}} \erfc(i t_{1-})   \right) +  \\
 \frac{1}{2w_{2}}\left( e^{-(t_{2+})^{2}} \erfc(it_{2+}) + e^{-(t_{2-})^{2}} \erfc(-i t_{2-})   \right),
\end{array}
\label{Eq:H2ErfcRepresentation}
\end{equation}
where
\begin{equation}
 t_{1\pm}=\frac{1}{2}(u_{1}+u_{2}\pm w_{1}), \quad t_{2\pm}= \frac{1}{2}(u_{1}+u_{2}\pm w_{2}),
 \label{Eq:t1t2roots}
\end{equation}
and where
\begin{equation}
 w_{1}=\sqrt{4ia+(u_{1}-u_{2})^{2}}, \quad w_{2}=\sqrt{-4ia+(u_{1}-u_{2})^{2}}.
 \label{Eq:w1w2}
\end{equation}
Representation for $a<0$ results from the reflection symmetry $H_{2}(-a,u_{1},u_{2}) = -H_{2}(a,u_{1},u_{2})$ of Proposition \ref{Proposition:reflection}.
\end{Proposition}
%%%%%%%%%%%%%%%%%%%%%%%%%%%%%%%%%%%%%%%%%%%%%%%%%

The next function that is important in application is the dumping 
function that shows how the maximum of 
(\ref{normalizedRelativisticVoigtProfile}) changes as a function of 
$\sigma$ - the smearing of a particle beam or a detector sensitivity. 
In applications, this function allows one to estimate the maximum of the 
convolution when a given smearing is selected. We define
\begin{equation}
 D_{0}(\sigma;\Gamma, \mu):= \frac{V_{0}(\mu;\mu,\Gamma,\sigma)}
 {\sigma_{0}(\mu;\mu,\Gamma)}
 \label{D_0},
\end{equation}
\begin{equation}
 D_{2}(\sigma;\Gamma, \mu):= \frac{V_{2} (\mu;\mu,\Gamma,\sigma)}
 {\sigma_{2}(\mu;\mu,\Gamma)}.
 \label{D_2}
\end{equation}
These functions are normalized in such a way that for $\Gamma \neq 0$ 
and $\mu \neq 0$
\begin{equation}
D_{0}(0;\Gamma, \mu)=D_{2}(0;\Gamma, \mu)=1,
\end{equation}
according to the (\ref{Normalization_lim_sigma0}) in Appendix \ref{Appendix_Normalization}. 
Plots of these functions are presented in Fig. \ref{Fig:Di}, 
\ref{Fig:DiMu} and \ref{Fig:DiGamma}.
%%%%%%%%%%%%%%%%%%%%%%%%%%%%%%%%%%%%%%%%%%%%%%%
\begin{figure}[!htb]
\centering
    \includegraphics[width=0.8\textwidth]{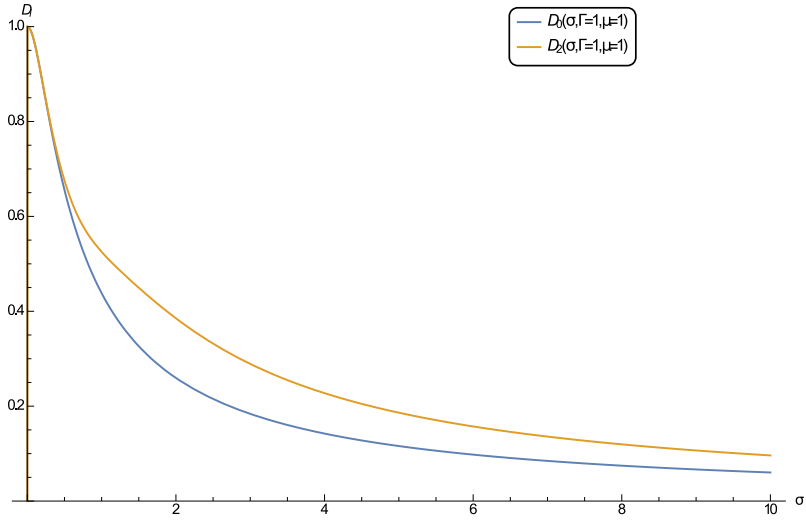}
    \caption{Dumping functions of (\ref{D_0}) and (\ref{D_2}).}
    \label{Fig:Di}
\end{figure}
%%%%%%%%%%%%%%%%%%%%%%%%%%%%%%%%%%%%%%%%%%%%%%%%
%%%%%%%%%%%%%%%%%%%%%%%%%%%%%%%%%%%%%%%%%%%%%%%
\begin{figure}[!htb]
\centering
    \includegraphics[width=0.8\textwidth]{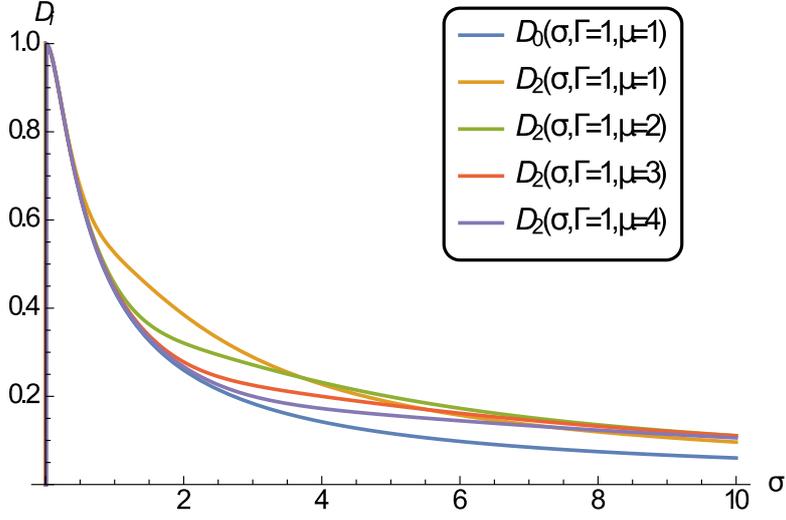}
    \caption{Dumping functions (\ref{D_0}) and (\ref{D_2}) for a few 
    values of $\mu$ parameter. It can be noted than when $\mu$ is 
    increased then the departure of the relativistic $D_{2}$ from 
    corresponding nonrelativistic $D_{0}$ is for higher values of 
    $\sigma$.}
    \label{Fig:DiMu}
\end{figure}
%%%%%%%%%%%%%%%%%%%%%%%%%%%%%%%%%%%%%%%%%%%%%%%%
%%%%%%%%%%%%%%%%%%%%%%%%%%%%%%%%%%%%%%%%%%%%%%%
\begin{figure}[!htb]
\centering
    \includegraphics[width=0.8\textwidth]{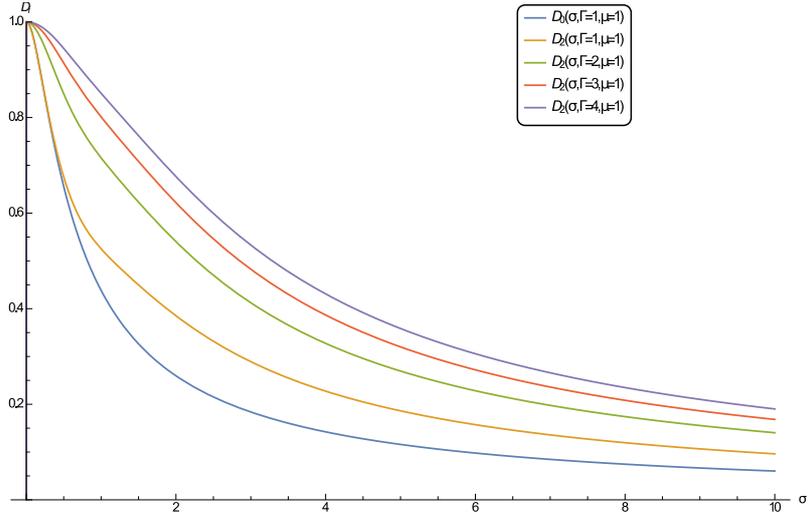}
    \caption{Dumping functions (\ref{D_0}) and (\ref{D_2}) for a few 
    values of $\Gamma$ parameter. As $\Gamma$ increases the 
    distribution $D_{2}$ becomes more flat around $\sigma =0$.}
    \label{Fig:DiGamma}
\end{figure}
%%%%%%%%%%%%%%%%%%%%%%%%%%%%%%%%%%%%%%%%%%%%%%%%

The relativistic line broadening function has an alternative representation
\begin{Proposition}
The relativistic line broadening function for $a>0$ can be presented as
\begin{equation}
 H_{2}(a,u_{1},u_{2})=\frac{1}{\pi}\int_{-\infty}^{\infty}dt e^{-t^{2}} \int_{0}^{\infty}e^{-ax}cos(x(t-u_{1})(t-u_{2}))dx
\end{equation}
or alternatively
\begin{equation}
\begin{array}{c}
 H_{2}(a,u_{1},u_{2}) =  \frac{1}{\pi}Re\int_{0}^{\infty}dx \frac{1}{\sqrt{\pi}\sqrt{1-ix}}e^{-ax+\frac{1}{4}ix\left(4u_{1}u_{2}-\frac{(u_{1}+u_{2})^{2}x}{i+x}\right)}.
\end{array}
\end{equation}
\end{Proposition}
The proof will be given in Appendix \ref{Appendix_Integral_representation}.

%%%%%%%%%%%%%%%%%%%%%%%%%%%%%%%%%%%%%%%%%%%%%%%%%
\section{Conclusions}
Definition of the relativistic Voigot profile and the line broadening function was provided 
along the same way as the original Voigt function. In addition, some properties and comparison of the dumping functions for both 
profiles was given.

We believe that this popular in use function will be included in modern tables of special functions.

%%%%%%%%%%%%%%%%%%%%%%%%%%%%%%%%%%%%%%%%%%%%%%%%
\section*{Acknowledgments}
RK research was supported by the GACR grant 17-19437S, and by the grants: MUNI/A/1103/2016 and MUNI/A/1138/2017 of Masaryk University. We would like to thanks the Anonymous Referee that helped us to reformulate results without use of the distribution theory, i.e., within the realm of special functions only.

\appendix

%%%%%%%%%%%%%%%%%%%%%%%%%%%%%%%%%%%%%%%%%%%%%%%%
\section{Appendix - Voigt profile}
\label{Appendix_Voigt}
%%%%%%%%%%%%%%%%%%%%%%%%%%%%%%%%%%%%%%%%%%%%%%%%

In this section for the sake of completeness the definition of normalized 
form of the Voigt distribution of (\ref{convolution}) for $i=0$ will be provided, as a comparison and motivation for our studies. 
The results are standard and details can be found in, e.g., 
\cite{AtomicResonances,H0table,V0function}.

The normalized relativistic Voigt can be obtained in the following way. Starting from
\begin{equation}
 V_{0}(E;\mu,\Gamma,\sigma)=\int_{-\infty}^{+\infty}\frac{\Gamma}
 {2\pi}\frac{1}{(E'-\mu)^{2}+\left(\frac{\Gamma}
 {2}\right)^{2}}\frac{1}{\sigma\sqrt{2\pi}}e^{-\frac{(E-E')^{2}}
 {2\sigma^{2}}} dE',
\end{equation}
upon introducing the new variables and constants
\begin{equation}
 t:=\frac{E-E'}{\sqrt{2}\sigma}, \quad a:=\frac{\Gamma}
 {2\sqrt{2}\sigma}, \quad u:=\frac{E-\mu}{\sqrt{2}\sigma},
\label{Voigt_variables}
\end{equation}
we obtain
\begin{equation}
V_{0}(E;\mu,\Gamma,\sigma)= \frac{1}{\sqrt{2\pi}\sigma}\frac{a}
{\pi}\int_{-\infty}^{\infty}\frac{e^{-t^{2}}}{(u-t)^{2}+a^{2}}dt.
\label{normalizedVoigtProfile}
\end{equation}
Defining new function (usually it is denoted by $H(a,u)$)
\begin{equation}
 H_{0}(a,u):=\frac{a}{\pi}\int_{-\infty}^{\infty}\frac{e^{-t^{2}}}
 {(u-t)^{2}+a^{2}}dt,
 \label{line_broadening_function}
\end{equation}
called line broadening function, the equation 
(\ref{normalizedVoigtProfile}) can be written in the form
\begin{equation}
 V_{0}(E;\mu,\Gamma,\sigma)= \frac{H_{0}(a,u)}{\sqrt{2\pi}\sigma},
\end{equation}
which is similar to (\ref{V2H2correspondence}) when considering the 
relativistic case.

% Using analogous as in the Appendix 
% \ref{Appendix_Normalization} and rather standard considerations, namely the Proposition 
% \ref{Proposition-multidelta}, one gets
% \begin{equation}
%  lim_{a\rightarrow 0^{\pm}} \frac{a}{\pi}\frac{1}{(t-u)^{2}+a^{2}} = 
%  \pm\delta(t-u),
% \end{equation}
% where $\delta$ is the Dirac delta distribution, and as the result

The limit $a\rightarrow 0^{\pm}$ can be derived using the complex generalization of the error function (see Chapter 7 of \cite{NIST})
\begin{equation}
 \begin{array}{c}
  w(z) = e^{-z^{2}}\erfc(-iz), \quad \forall z, \\
  w(z) = \frac{1}{i\pi}\int_{-\infty}^{\infty}\frac{e^{-t^{2}}}{t-z}dt, \quad \Im z >0,
 \end{array}
 \label{Eq:erfcRepresentation}
\end{equation}
where $\erfc(z)$ is the complementary error function.

Factoring (\ref{line_broadening_function}) into the sum of simple rational terms for $a>0$ one gets
\begin{equation}
 H_{0}(a,u)=\frac{1}{2}\left( e^{-(u+ia)^{2}} \erfc(a-iu)+e^{-(u-ia)^{2}} \erfc(a+iu) \right),
\end{equation}
and for $a<0$ the similar expression
\begin{equation}
H_{0}(a,u)=-\frac{1}{2}\left( e^{-(u+ia)^{2}}\erfc(-a+iu)+e^{-(u-ia)^{2}}\erfc(-a-iu) \right).
\end{equation}
Passing to the limit $a\rightarrow 0^{\pm}$ and using the property $\erfc(z) + \erfc(-z)=2$ standard result is obtained
\begin{equation}
 lim_{a\rightarrow 0^{\pm}} H_{0}(a,u) = \pm e^{-u^{2}}.
 \label{Appendix_Voigt_H0_a0plusminus}
\end{equation}
One can note that when comparing it with the relativistic version 
(\ref{H2_limit_a_0}) there is no singularity here at $a=0$ (as for 
$u_{1}=u_{2}$ in the relativistic case).
%%%%
Line broadening functions are presented in Fig. \ref{Fig:H0}.
%%%%%%%%%%%%%%%%%%%%%%%%%%%%%%%%%%%%%%%%%%%%%%%
\begin{figure}[!htb]
\centering
    \includegraphics[width=0.8\textwidth]{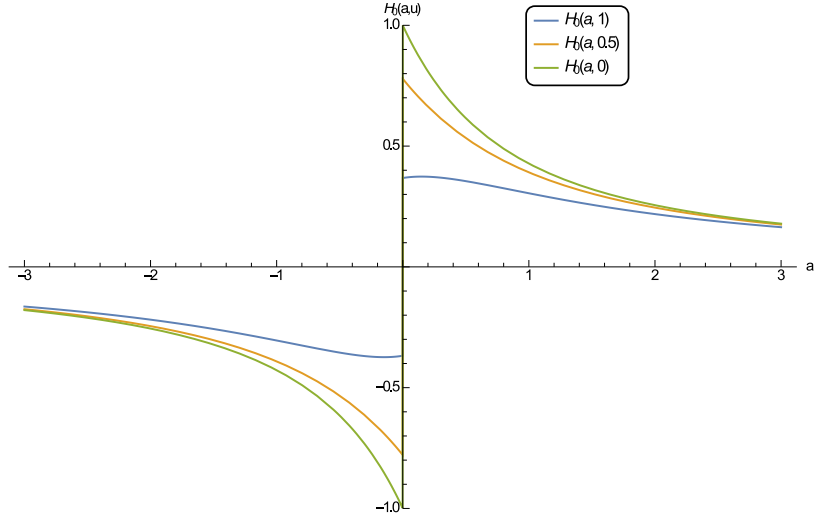}
    \caption{Voigt distribution (\ref{line_broadening_function}) for 
    a few values of parameters. The limiting values at $a \rightarrow 
    0^{\pm}$ is given by (\ref{Appendix_Voigt_H0_a0plusminus}). 
    Compare it with \ref{Fig:H2}.}
    \label{Fig:H0}
\end{figure}
%%%%%%%%%%%%%%%%%%%%%%%%%%%%%%%%%%%%%%%%%%%%%%%%

We can also provide alternative representation for 
(\ref{line_broadening_function}) using the following Proposition from
\cite{AtomicResonances}, which will be generalized to the relativistic Voigt profile in the Appendix \ref{Appendix_Integral_representation}
\begin{Proposition}
 \begin{equation}
  \frac{a}{(t-u)^{2}+a^{2}}=\int_{0}^{\infty}e^{-ax}\cos((t-u)x)dx
 \end{equation}
 for $a>0$.
\end{Proposition}
The proof is as follows
\begin{equation}
\begin{array}{c}
\int_{0}^{\infty}e^{-ax}\cos((t-u)x)dx=Re\int_{0}^{\infty} e^{-
ax}e^{i((t-u)}dx =  \\
=Re \frac{-1}{-a+i(t-u)} =  \frac{a}{(t-u)^{2}+a^{2}}.
\end{array}
\end{equation}
This allows us to rewrite (\ref{line_broadening_function}) in the 
following way
\begin{equation}
 \begin{array}{c}
  H_{0}(a,u)=\frac{1}{\pi}\int_{-\infty}^{\infty}dt e^{-
  t^{2}}\int_{0}^{\infty}dx e^{-ax}\cos((t-u)x) = \\
  \frac{1}{\pi}\int_{0}^{\infty}dx e^{-ax}\int_{-\infty}^{\infty}dt 
  e^{-t^{2}}(\cos(ux)\cos(ut)+\sin(ux)\sin(xt)) = \\
  =\frac{1}{\pi}\int_{0}^{\infty}dx e^{-
  ax}cos(ux)\int_{-\infty}^{\infty}dt e^{-t^{2}}\cos(xt) = \\
  \frac{1}{\sqrt{\pi}}\int_{0}^{\infty}dx e^{-ax-\frac{x^{2}}
  {4}}\cos(ux).
 \end{array}
\end{equation}
Switching to complex representation we finally obtain
\begin{equation}
 H_{0}(a,u)=Re \frac{1}{\sqrt{\pi}}\int_{0}^{\infty}dx e^{-
 ax+iux-\frac{x^{2}}{4}}.
\end{equation}
This can be converted to the Faddeeva function described in \cite{NIST}.

%%%%%%%%%%%%%%%%%%%%%%%%%%%%%%%%%%%%%%%%%%%%%%%%%%%%%%%%%%%
\section{Appendix - Limiting value for $a \rightarrow 0$}
\label{Appendix_Normalization}
%%%%%%%%%%%%%%%%%%%%%%%%%%%%%%%%%%%%%%%%%%%%%%%%%%%%%%%%%%%

In this section the normalization of (\ref{relativistic_line_broadening_function}) in the limit 
$a\rightarrow 0$ will be derived. Define
\begin{equation}
 I_{2}(u_{1},u_{2},a):=\frac{a}{\pi}\int_{-\infty}^{\infty}\frac{dt}
 {(t-u_{1})^{2}(t-u_{2})^{2}+a^{2}}.
 \label{Appendix_Normalization_I2}
\end{equation}
As an introduction consider the following
\begin{Proposition}
For $u_{1}\neq u_{2}$ we have
 \begin{equation}
 \begin{array}{c}
  lim_{a\rightarrow 0^{+}} I_{2}(u_{1},u_{2},a) = \frac{2}{|u_{1}-
  u_{2}|}  \\
  lim_{a\rightarrow 0^{-}} I_{2}(u_{1},u_{2},a) = \frac{-2}{|u_{1}-
  u_{2}|}  
 \end{array}
 \label{Appendix_Normalization_I2_limit}
 \end{equation}
\end{Proposition}

The proof is simple application or the Cauchy residue theorem 
\cite{AblowitzFokas}. The zeros of the integrand denominator are 
located at
\begin{equation}
 t_{1\pm}=\frac{1}{2}(u_{1}+u_{2}\pm w_{1}), \quad t_{2\pm}= \frac{1}{2}(u_{1}+u_{2}\pm w_{2}),
 \label{Eq:t1t2roots}
\end{equation}
where
\begin{equation}
 w_{1}=\sqrt{4ia+(u_{1}-u_{2})^{2}}, \quad w_{2}=\sqrt{-4ia+(u_{1}-u_{2})^{2}}.
 \label{Eq:w1w2}
\end{equation}

Assume that $a>0$ and consider the contour of integration in the 
complex plane presented in Fig. \ref{Fig:Contour}.
%%%%%%%%%%%%%%%%%%%%%%%%%%%%%%%%%%%%%%%%%%%%%%%
\begin{figure}[!htb]
\centering
    \includegraphics[width=0.8\textwidth]{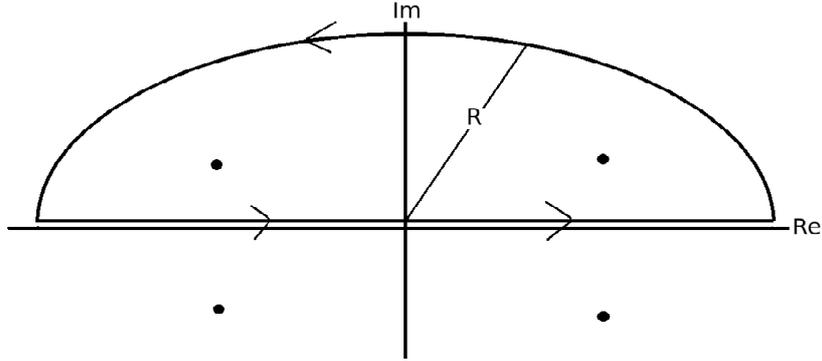}
    \caption{Countour of integration in the complex plane for (\ref{Appendix_Normalization_I2}). The dots show the poles and $R \rightarrow \infty$.}
    \label{Fig:Contour}
\end{figure}
%%%%%%%%%%%%%%%%%%%%%%%%%%%%%%%%%%%%%%%%%%%%%%%%
We have
\begin{equation}
\begin{array}{c}
 \int_{-R}^{R} \frac{a}{\pi}\frac{dt}{(t-u_{1})^{2}(t-
 u_{2})^{2}+a^{2}} + \int_{\smallfrown} \frac{a}{\pi}\frac{dt}{(t-
 u_{1})^{2}(t-u_{2})^{2}+a^{2}} = \\ \\
 2\pi i \sum_{i} res_{i}\frac{a}{\pi}\frac{1}{(t-u_{1})^{2}(t-
 u_{2})^{2}+a^{2}},
\end{array}
\end{equation}
where the second integral is over the half circle of the radius $R$ 
in the upper complex half-plane and the sum is over the residues 
inside the contour, i.e., at $t_{1+}$ and  at $t_{2-}$. Noting that in the limit $R \rightarrow \infty$ the 
integral over the semicircle vanishes one has from the Cauchy 
residue theorem that
\begin{equation}
\begin{array}{c}
 I_{2}(u_{1},u_{2},a) = \frac{1}{w1} + \frac{1}{w2}.
\end{array}
\end{equation}
Taking the limit $a \rightarrow 0^{+}$ one gets the first result of 
(\ref{Appendix_Normalization_I2_limit}). The same result can be 
obtained if the circle in the lower complex semiplane is used.

In order to prove the second statement it should be assumed that 
$a<0$. Then the poles in the contour are interchanged
with those that are outside the contour and as a result it generates 
additional minus sign. The second equation of (\ref{Appendix_Normalization_I2_limit}) can be also deduced from the first one and the fact that $I(u_{1},u_{2},-a)=-I(u_{1},u_{2},a)$. 
This ends the proof.

It is worth noting that from the above reasoning using the same 
contour of integration we have
\begin{Corollary}
 For an analytic function with no poles in the upper complex half-plane $\phi(t)$ 
 such that $\lim_{|t|\rightarrow\infty} |\phi(t)| = 0$ we have
 \begin{equation}
  \lim_{a\rightarrow 0^{\pm}} \int_{-\infty}^{\infty} \frac{a}
  {\pi}\frac{\phi(t)}{(t-u_{1})^{2}(t-u_{2})^{2}+a^{2}}dt = \frac{\pm 
  1}{|u_{1}-u_{2}|}(\phi(u_{1})+\phi(u_{2})),
 \end{equation}
 when $u_{1}\neq u_{2}$.
\end{Corollary}

The Corollary cannot be used for $\phi(t)$ being the Gaussian function 
(\ref{GaussianPDF}) as it is unbounded for $Im(t)\rightarrow 
\pm\infty$, however, the behaviour from the Corollary is also valid 
for the Gaussian distribution as it will be shown below. 

% The obstacle is as follows, if one would like to use in the proof the technique of the Cauchy residue theorem it would require some complicated 
% contour of integration that encloses the poles of (\ref{Appendix_Normalization_I2}) and does not contain the sector of $Arg(t) \in \left[ \frac{\pi}{4}, \frac{3\pi}{4} \right]$. 

In order to overcome these difficulties the contour of integration will be modified in the following way. Assuming that $a>0$ and selecting the rectangle contour which consists the real line and a parallel horizontal line that has $\Im t$ greater than both $\Im t_{1+}$ and $\Im t_{2-}$ (see Fig. \ref{Fig:ContourRectangle}) the auxiliary formula of the integral over the contour can be rewritten as the sum of the four integrals
\begin{equation}
\begin{array}{c}
 \left( \int_{-R}^{R} + \int_{\uparrow} + \int_{\mathcal{L}}  + \int_{\downarrow}  \right)  \frac{a}{\pi}\frac{e^{-t^{2}}dt}{(t-u_{1})^{2}(t-u_{2})^{2}+a^{2}} = \\ \\
 2\pi i \sum_{\bar{t} \in \{t_{1+},t_{2-}\}} res_{t=\bar{t}} \frac{a}{\pi}\frac{e^{-t^{2}}}{(t-u_{1})^{2}(t-  u_{2})^{2}+a^{2}},
\end{array}
\end{equation}
where the arrows denote the integration along the vertical sides of the rectangle.
%%%%%%%%%%%%%%%%%%%%%%%%%%%%%%%%%%%%%%%%%%%%%%%
\begin{figure}[!htb]
\centering
    \includegraphics[width=0.8\textwidth]{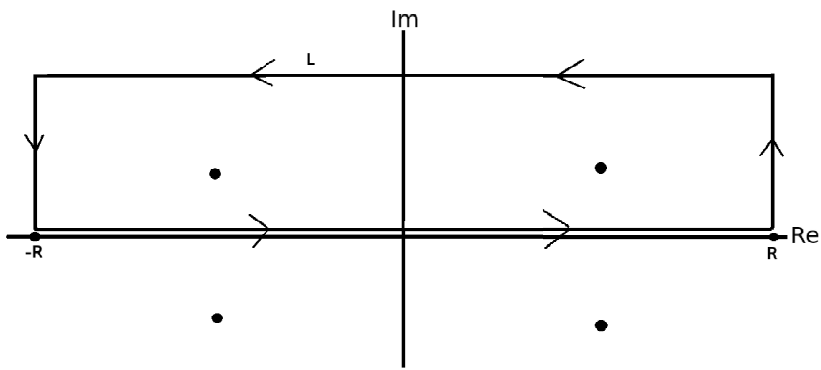}
    \caption{Countour of integration in the complex plane for (\ref{relativistic_line_broadening_function}). The dots show the poles and $\mathcal{L}$ line is parallel to the real line.}
    \label{Fig:ContourRectangle}
\end{figure}
%%%%%%%%%%%%%%%%%%%%%%%%%%%%%%%%%%%%%%%%%%%%%%%%
Passing to the limit $R \rightarrow \infty$ and noting that the integrals over vertical segments vanishes, the formula (\ref{relativistic_line_broadening_function}) can be written as
\begin{equation}
H_{2}(a, u_{1}, u_{2}) = \frac{a}{\pi} \int_{\mathcal{L}}\frac{e^{-t^{2}}dt}{(t-u_{1})^{2}(t-u_{2})^{2}+a^{2}} + \frac{e^{-t_{1+}^{2}}}{w_{1}} + \frac{e^{-t_{2-}^{2}}}{w_{2}}.
\label{Eq:H2RectangleExpression}
\end{equation}
This form can be used to derive the limit $a\rightarrow 0^{+}$: The integral over $\mathcal{L}$ vanishes as the integrand is bounded on $\mathcal{L}$ and because of the factor $a$ in front of the integral. The limits of (\ref{Eq:t1t2roots}) and (\ref{Eq:w1w2}) are also easily computed by setting $a=0$.

By analogous reasoning for $a<0$ or using reflection symmetry from the Proposition \ref{Proposition:reflection}, i.e. $H_{2}(-a,u_{1},u_{2})=-H_{2}(a,u_{1},u_{2})$, the other limit $a \rightarrow 0^{-}$ can also be computed.

From this easily follows that
\begin{equation}
 lim_{a\rightarrow 0^{\pm}}H_{2}(a,u_{1},u_{2})=\frac{\pm}{|u_{1}-
 u_{2}|}(e^{-u_{1}^{2}}+e^{-u_{2}^{2}}).
 \label{Appendix_Normalization_H2_a0}
\end{equation}
These results give the limit for (\ref{convolution}), the case $i=2$, 
in the form
\begin{equation}
 lim_{\Gamma\rightarrow 0^{\pm}} V_{2}(E;\sigma,\mu,\Gamma) 
 =\frac{\pm}{\sigma\mu2\sqrt{2\pi}}\left( e^{-\frac{(E-\mu)^{2}}
 {2\sigma^{2}} } + e^{-\frac{(E+\mu)^{2}}{2\sigma^{2}} } \right).
\end{equation}

% The Proposition \ref{Proposition-multidelta} for $N=1$ can be also 
% used for examining the limit $\sigma \rightarrow 0$ for 
% (\ref{GaussianPDF}), namely, the Gaussian distribution converges to 
% the Dirac delta distribution, which is the well known fact(see, e.g., 
% \cite{DeltaSeries, DistributionZemanian, DistributionMikusinski})
% \begin{equation}
%  \lim_{\sigma\rightarrow 0} G(E-E_{0};\sigma)=\delta(E-E_{0}).
% \end{equation}

We can also express $H_{2}$ in terms of the complementary error function as in the Appendix \ref{Appendix_Voigt} for $H_{0}$. Firstly, the equation (\ref{relativistic_line_broadening_function}) can be transformed to the form
\begin{equation}
 \begin{array}{c}
H_{2}(a, u_{1}, u_{2}) = \frac{1}{2\pi i}\int_{-\infty}^{\infty}dt \left( \frac{1}{w_{1}} \left( \frac{e^{-t^{2}}}{t-t_{1+} } - \frac{e^{-t^{2}}}{t-t_{1-} } \right) - \frac{1}{w_{2}} \left( \frac{e^{-t^{2}}}{t-t_{2+} } - \frac{e^{-t^{2}}}{t-t_{2-} } \right) \right).
 \end{array}
\end{equation}
Then, for the case $a>0$, using (\ref{Eq:erfcRepresentation}) it results that
\begin{equation}
\begin{array}{c}
 H_{2}(a, u_{1}, u_{2}) = \frac{1}{2w_{1}}\left( e^{-(t_{1+})^{2}} \erfc(-it_{1+}) + e^{-(t_{1-})^{2}} \erfc(i t_{1-})   \right) +  \\
 \frac{1}{2w_{2}}\left( e^{-(t_{2+})^{2}} \erfc(it_{2+}) + e^{-(t_{2-})^{2}} \erfc(-i t_{2-})   \right), 
\end{array}
\end{equation}
which restores the results of Proposition \ref{Proposition:H2ErfcRepresentation}. Similar representation for $a<0$ is derived using the reflection symmetry $H_{2}(u_{1},u_{2},-a) = -H_{2}(u_{1},u_{2},a)$. Finally, passing to the limit $a\rightarrow 0^{\pm}$ and using, as in Appendix \ref{Appendix_Voigt}, the property $\erfc(z)+\erfc(-z)=2$, the limit (\ref{Appendix_Normalization_H2_a0}) is restored.

Similar rectangle contour can be used for examining the limit $\sigma \rightarrow 0$ for (\ref{GaussianPDF}) and as a rather standard result we have
\begin{equation}
 \lim_{\sigma\rightarrow 0}V_{2}(E;\mu,\Gamma,\sigma)= \sigma_{2}(E;\mu,\Gamma).
 \label{Normalization_lim_sigma0}
\end{equation}

In the last part we examine 
(\ref{relativistic_line_broadening_function}) on the manifold 
$u_{1}=u_{2}=u$. Under this constraint the following integral 
\begin{equation}
 H_{2}(a,u,u)= \frac{a}{\pi}\int_{-\infty}^{\infty} 
 \frac{e^{-t^{2}}}{(u-t)^{4}+a^{2}} dt
 \label{Appendix_Normalization_H2_u1=u2}
\end{equation}
is bounded for $a \neq 0$ as the integrand is a product of the bounded function 
$e^{-t^{2}}$ and the integrable function. When $a\rightarrow 0^{+}$ 
then the integrand function has nonintegrable singularity\footnote{The case $u_{1}=u_{2}$ 
for $a\neq0$ corresponds to $\mu=0$ according to (\ref{Voigt_variables}), i.e., in particle physics to a massles particle of some nonzero width $\Gamma$. Only when $\Gamma\rightarrow 0$ then $V_{2}$ becomes unbounded. It is assumed that the spread $\sigma >0$.} at $t=u_{1}$ and the integral tends to $\infty$. When $a \rightarrow 0^{-}$ then 
(\ref{Appendix_Normalization_H2_u1=u2}) tends to $-\infty$, which results from antisymmetry with respect to $a$. Explicit singularity asymptotics can be obtained using the Laurent series expansion of (\ref{Eq:H2RectangleExpression}) for $a\approx 0$
\begin{equation}
 H_{2}(a,u_{1}=u,u_{2}=u)=\frac{e^{-u^2}}{\sqrt{2a}}+ \frac{e^{-u^{2}}}{\sqrt{2}}(2u^{2}-1)\sqrt{a} + O(a),
\end{equation}
where the integral over $\mathcal{L}$ is contained in the $O(a)$ term. The same expansion can be obtained from the Laurent series expansion of (\ref{Eq:H2ErfcRepresentation}).

%%%%%%%%%%%%%%%%%%%%%%%%%%%%%%%%%%%%%%%%%%%%%%%%
\section{Appendix - Limiting values for $u_{1},u_{2} \rightarrow \infty$, $a\neq0$}
\label{Appendix_u1u2large}
%%%%%%%%%%%%%%%%%%%%%%%%%%%%%%%%%%%%%%%%%%%%%%%%
In this section an approximate formula for large 
$u_{1}$, $u_{2}$ and $a\neq 0$ will be provided. Under these 
assumption the main contribution to the integral 
(\ref{relativistic_line_broadening_function}) is from the vicinity of 
$t=0$ point. Expanding the following function around $t=0$ one gets
\begin{equation}
 \frac{a}{\pi}\frac{1}{(t-u_{1})^{2}(t-u_{2})^{2}+a^{2}}=\frac{a}
 {\pi}\frac{1}{u_{1}^{2}u_{2}^{2}+a^{2}}+O(t),
\end{equation}
which is uniformly convergent in some closed interval around $t=0$. 
This interval can be made large enough(when $|u_{1}|$ and $|u_{2}|$ 
are large) that the Gaussian integral over it is close to the 
integral over the whole real line. Then it can be integrated term by 
term and as the result one gets approximation of the form
\begin{equation}
 H_{2}(u_{1},u_{2},a)\approx \frac{a}{\sqrt{\pi}
 (u_{1}^{2}u_{2}^{2}+a^{2})}+\ldots,
\end{equation}
where other terms vanishes faster than the first term as $|u_{1}|$ and $|u_{2}|$ tends to infinity.

The similar approach can be employed in the case of (\ref{V2}) for 
$\sigma \ll \Gamma$ at $E=\mu$. Under these assumptions the 
relativistic Breit-Wigner distribution under the integral can be 
expanded around $E=\mu$. One gets
\begin{equation}
 V_{2}(\mu;\mu,\Gamma,\sigma)\approx \sigma_{2}(\mu;\mu,
 \Gamma)+\ldots,
\end{equation}
where remaining terms vanishes for $\sigma \rightarrow 0$. This result restores the property of (\ref{D_2}).

%%%%%%%%%%%%%%%%%%%%%%%%%%%%%%%%%%%%%%%%%%%%%%%%
\section{Appendix - Integral representation}
\label{Appendix_Integral_representation}
%%%%%%%%%%%%%%%%%%%%%%%%%%%%%%%%%%%%%%%%%%%%%%%%
In this section we provide another integral representation of $H_{2}$
function similar to those in Appendix \ref{Appendix_Voigt}.

First we provide transformation of the rational part under the assumption $a>0$, i.e., 
\begin{equation}
\begin{array}{c}
 \frac{1}{(t-u_{1})^{2}(t-u_{2})^{2}+a^{2}} =\frac{-1}{a} Re\frac{-a-i(t-u_{1})(t-u_{2})}{(t-u_{1})^{2}(t-u_{2})^{2}+a^{2}} =  \\ \\
 \frac{-1}{a}Re\frac{1}{-a+i(t-u_{1})(t-u_{2})} = \\ \\
 \frac{1}{a}Re\int_{0}^{\infty}e^{x(-a+i(t-u_{1})(t-u_{2}))}dt = \\ \\
 \frac{1}{a}\int_{0}^{\infty} e^{-ax}\cos(x(t-u_{1})(t-u_{2})) dt.
\end{array}
\end{equation}
Introducing this into $H_{2}$ we obtain double integral representation
\begin{equation}
 H_{2}(a,u_{1},u_{2})=\frac{1}{\pi}\int_{-\infty}^{\infty}dt e^{-t^{2}} \int_{0}^{\infty}e^{-ax}cos(x(t-u_{1})(t-u_{2}))dx
\end{equation}
or alternatively
\begin{equation}
\begin{array}{c}
 H_{2}(a,u_{1},u_{2})=\frac{1}{\pi}Re \int_{0}^{\infty}dt e^{-ax} \int_{-\infty}^{\infty}e^{-t^{2}+ix(t-u_{1})(t-u_{2}))}dx = \\ \\
 \frac{1}{\pi}Re\int_{0}^{\infty}dx \frac{1}{\sqrt{\pi}\sqrt{1-ix}}e^{-ax+\frac{1}{4}ix\left(4u_{1}u_{2}-\frac{(u_{1}+u_{2})^{2}x}{i+x}\right)}.
\end{array}
\end{equation}
These expressions is, however, more complicated comparing to the original one for $H_{2}$, i.e., equation (\ref{relativistic_line_broadening_function}).

%%%%%%%%%%%%%%%%%%%

\end{document}